# Enhancing mTBI Diagnosis with Residual Triplet Convolutional Neural Network Using 3D CT

Hanem Ellethy,[1] Shekhar S. Chandra[1] and Viktor Vegh[2,3]

[1] School of Electrical Engineering and Computer Science, University of Queensland, QLD, Australia.

[2] The Centre for Advanced Imaging, Australian Institute for Bioengineering and Nanotechnology, University of Queensland, QLD, Australia.

[3] ARC Training Centre for Innovation in Biomedical Imaging Technology, QLD, Australia.

*Correspondence: Hanem Ellethy, email: h.elwaseif@uq.edu.au, phone: +61423403254, address: School of Electrical Engineering and Computer Science, GP Building 78, The University of Queensland, St Lucia QLD 4072, Australia.

## Abstract

Mild Traumatic Brain Injury (mTBI) is a common and challenging condition to diagnose accurately. Timely and precise diagnosis is essential for effective treatment and improved patient outcomes. Traditional diagnostic methods for mTBI often have limitations in terms of accuracy and sensitivity. In this study, we introduce an innovative approach to enhance mTBI diagnosis using 3D Computed Tomography (CT) images and a metric learning technique trained with triplet loss. To address these challenges, we propose a Residual Triplet Convolutional Neural Network (RTCNN) model to distinguish between mTBI cases and healthy ones by embedding 3D CT scans into a feature space. The triplet loss function maximizes the margin between similar and dissimilar image pairs, optimizing feature representations. This facilitates better context placement of individual cases, aids informed decision-making, and has the potential to improve patient outcomes. Our RTCNN model shows promising performance in mTBI diagnosis, achieving an average accuracy of 94.3%, a sensitivity of 94.1%, and a specificity of 95.2%, as confirmed through a five-fold cross-validation. Importantly, when compared to the conventional Residual Convolutional Neural Network (RCNN) model, the RTCNN exhibits a significant improvement, showcasing a remarkable 22.5% increase in specificity, a notable 16.2% boost in accuracy, and an 11.3% enhancement in sensitivity. Moreover, RTCNN requires lower memory resources, making it not only highly effective but also resource-efficient in minimizing false positives while maximizing its diagnostic accuracy in distinguishing normal CT scans from mTBI cases. The quantitative performance metrics provided and utilization of occlusion sensitivity maps to visually explain the model's decision-making process further enhance the interpretability and transparency of our approach.

**KEYWORDS**: Mild Traumatic Brain Injury, mTBI Diagnosis, metric learning, Triplet Loss, 3D CT Imaging, Deep Learning, Occlusion sensitivity maps (OSM)

# 1 Introduction

Mild Traumatic Brain Injury (mTBI) represents a pervasive and complex public health concern [1]. With its potential to cause short-term and long-lasting cognitive impairments, mTBI necessitates accurate and timely diagnosis for optimal patient care. Traditional diagnostic approaches for mTBI, including clinical evaluation and neuroimaging, are vital but may lack the sensitivity to detect subtle but clinically significant changes associated with mTBI [2], [3]. This limitation underscores the need for advanced imaging and analysis techniques to enhance diagnostic accuracy.

Computed Tomography (CT) stands as the primary neuroimaging modality for assessing traumatic brain injury (TBI) due to its widespread availability, cost-effectiveness, and fast screening [4]. Despite its limitations in detecting subtle or absent structural changes in cases of mTBI [5], CT plays a crucial role in guiding the treatment of acute TBI [6]. In contrast to conventional 2D imaging, 3D CT scans yield comprehensive volumetric data, facilitating the visualization of even subtle brain abnormalities and thereby enhancing diagnostic precision. This capability holds particular significance in mTBI diagnosis, where traditional 2D slices may not clearly reveal the exact location and extent of injury [7].

The emergence of deep learning (DL) has led to a substantial revolution in the field of medical image analysis. Convolutional Neural Networks (CNNs) have demonstrated remarkable achievements in diverse medical imaging tasks, encompassing image classification, segmentation, and feature extraction [8]. In the context of mTBI diagnosis, DL models have the potential to automatically extract highly discriminative features from 3D CT scans, aiding in the detection of subtle abnormalities. Nevertheless, the application of DL for diagnosing mTBI using CT data is not widely explored [9]. One noteworthy DL architecture that has demonstrated exceptional capabilities in image feature extraction is the Residual CNN (RCNN). The RCNN employs residual blocks to capture complex patterns and representations within medical images, making it a suitable choice for mTBI diagnosis [10].

Metric learning is a subfield of machine learning that focuses on optimizing feature representations for similarity-based tasks [11]. By leveraging metric learning techniques, researchers aim to enhance the separability of data points in feature space, making it easier to discriminate between different classes or categories [12]. In the context of medical image analysis, metric learning has gained interest because of its potential to improve the quality of extracted features and subsequently enhance diagnostic accuracy [13]–[16]. Metric learning presents a new approach for improving mTBI diagnosis, particularly when coupled with RCNN architectures, by offering extraction of discriminative and clinically relevant features from medical 3D scans [17].

One prominent approach within metric learning is the use of triplet loss (TL) functions. TL focuses on improving model discriminative abilities by minimizing the distance between similar data points while maximizing the distance between dissimilar ones in feature space. This approach encourages the deep learning model to focus on extracting features that are relevant for distinguishing between different classes of data.

This manuscript explores the utilization of metric learning with TL as an innovative approach to leverage the potential of 3D CT scans for enhancing mTBI diagnosis. We examine the impact of integrating metric learning into the RCNN model, which functions as an embedder, for the extraction of highly informative features from 3D CT images. Through a comparative analysis of the metric learning-integrated model against a conventional RCNN, our objective is to identify the diagnostic capabilities facilitated by this novel approach.

This study's contributions reach beyond the technical domain and have the potential to influence clinical practice by offering enhanced tools for mTBI diagnosis. Our proposed method, capable of detecting subtle brain abnormalities and extracting pertinent features, may expedite and improve the precision of mTBI diagnosis, ultimately contributing to enhanced patient outcomes. The main contribution of this study can be summarized as follows:

- We introduce the Residual Triplet Convolutional Neural Network (RTCNN) as an innovative tool for acute mTBI diagnosis using 3D CT imaging. Our approach integrates metric learning to optimize feature representations, enabling better context placement of individual cases, aiding informed decision-making, and potentially improving patient outcomes.

- We demonstrate the substantial impact of incorporating triplet loss, resulting in a remarkable 16.2% improvement in accuracy, an 11.3% increase in sensitivity, and a notable 22.5% enhancement in specificity, all achieved with more efficient utilization of memory resources.
- We utilize the Occlusion Sensitivity Maps (OSM) to improve RTCNN model interpretability by highlighting important image regions for decision-making, aiding in debugging, validating, and assessing trustworthiness.

## 2  Methods

### 2.1  Study Population

The study population for this research comprised individuals enrolled in the TRACK-TBI pilot study. TRACK-TBI is a comprehensive, multi-centre initiative aimed at investigating various aspects of TBI. TRACK-TBI pilot study is a prospective multi centre cohort study that enrolled 600 patients, older than seven years old, who suffered TBI. These patients underwent head CT scans in an emergency department within 24 hours of their head injuries. The study was conducted at three level I trauma centres: the University of California San Francisco, University Medical Centre Brackenridge, and the University of Pittsburgh Medical Centre, as well as the Mount Sinai Rehabilitation Centre, from April 2010 to June 2012 in the USA [18]. The TRACK-TBI dataset comprises demographic clinical data and CT images within 24 hours of head injury, MRI images within 14 days of head injury, and outcome assessment at three and six months. The TRACK-TBI pilot study was conducted to validate the feasibility of TBI common data elements (TBI-CDEs) implementation. The TRACK-TBI database contained a total of 589 records at the time of our access and analysis, out of the 600 subjects initially recruited.  Among these, 323 subjects had mTBI abnormalities on CT.

We were granted access rights to the FITBIR data repository by the Data Access and Quality (DAQ) committee. Our research, which utilized this dataset, received approval from the University of Queensland's human research ethics committee (no. 2020002583). Our study focused on patients who had received non-contrasted CT scans within 48 hours after experiencing closed head injuries and were assigned a Glasgow Coma Scale (GCS) score of ≥ 13. This resulted in a total of 296 de-identified records from the publicly available TRACK-TBI dataset.

### 2.2  Data Pre-processing

We conducted pre-processing on the CT images to ensure that high-quality and consistent data were fed to the various DL implementations [10]. First, we used the dcm2nii tool [19] to convert the CT images from their original format to the Neuroimaging Informatics Technology Initiative (NIfTI) format. This conversion included broad field-of-view alignment to ensure spatial consistency across all images. Next, we applied the appropriate brain window settings (window width = 80 Hounsfield Units (HU) and width level = 40 HU) [20] to standardize the dynamic range of the CT images and enhance the visibility of potential lesions associated with mTBI. To achieve isotropic resolution, we employed the Spline Interpolated Zoom (SIZ) resampling volume algorithm [21], resulting in images with a matrix size of 128×128×64 and a resolution of 1mm$^3$. This resizing and standardization process aimed to reduce computational complexity and optimize GPU usage. We used z-axis interpolation to ensure a uniform volume representation, effectively capturing data from multiple slices while adhering to GPU memory constraints [22]. Lastly, we removed unnecessary background information from the images, such as the scanner bed, to focus the model exclusively on the brain structures of interest, as depicted in Figure 1.

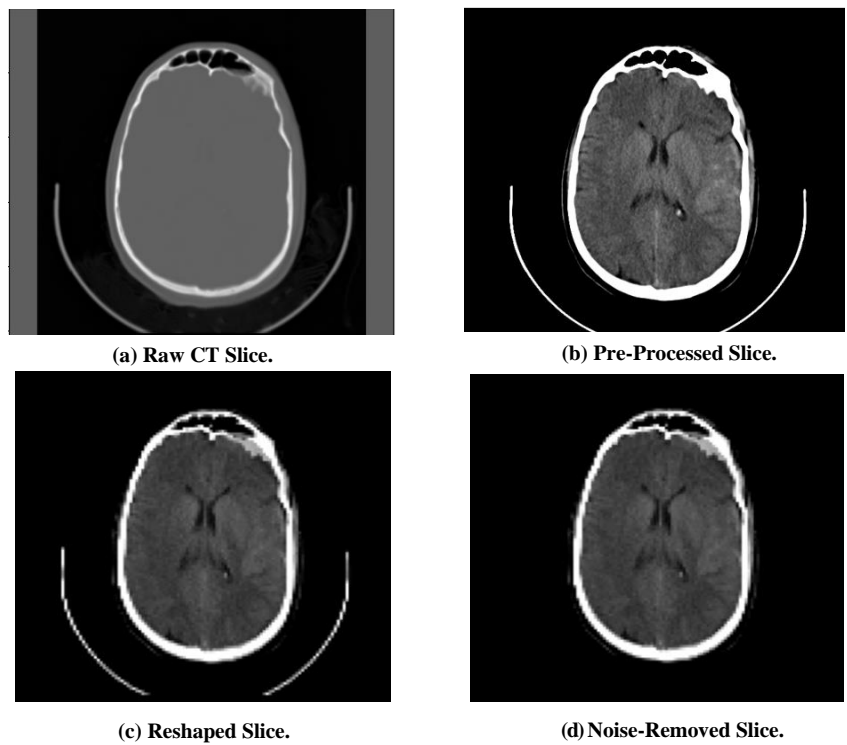

(a) Raw CT Slice.    (b) Pre-Processed Slice.

(c) Reshaped Slice.    (d) Noise-Removed Slice.

**Figure 1. The CT slice pre-processing steps.** (a) The original, unprocessed CT slice before any pre-processing steps were applied. (b) The same CT slice after undergoing initial pre-processing, which includes standardizing the dynamic range to enhance the visibility of brain anatomical structures. (c) The CT slice after being reshaped, which is a crucial part of the pre-processing steps. (d) The CT slice has been further processed to remove high-intensity noise artifacts such as the scanner bed.

## 2.3 Study design

In this study, we investigate how a metric learning technique, applied to a modified RCNN model [10], impacts the accuracy of mTBI diagnosis by enhancing its ability to detect subtle patterns in medical imaging data. First, our baseline RCNN model is employed to diagnose mTBI. Next, the RCNN architecture is integrated with triplet loss to assess the benefits of incorporating metric learning for more reliable mTBI diagnoses using CT images. Then, the performance of both models is evaluated along with visual assessment through the Occlusion Sensitivity Maps (OSM) [23].

### 2.3.1 RCNN model

The baseline model architecture RCNN is derived from our previous study [10]. The model is composed of two main components: an encoder and a classifier. Within the encoder, there are six consecutive residual blocks specifically structured to gradually reduce the dimensions of the 3D input image, as shown in Figure 2. This step is intended to capture the 3D feature representation specific to mTBI. Subsequently, the output from the encoder is flattened and channelled into the classifier. The classifier is composed of four densely connected layers, responsible for categorizing the extracted features as either indicative of normal or mTBI cases [10].

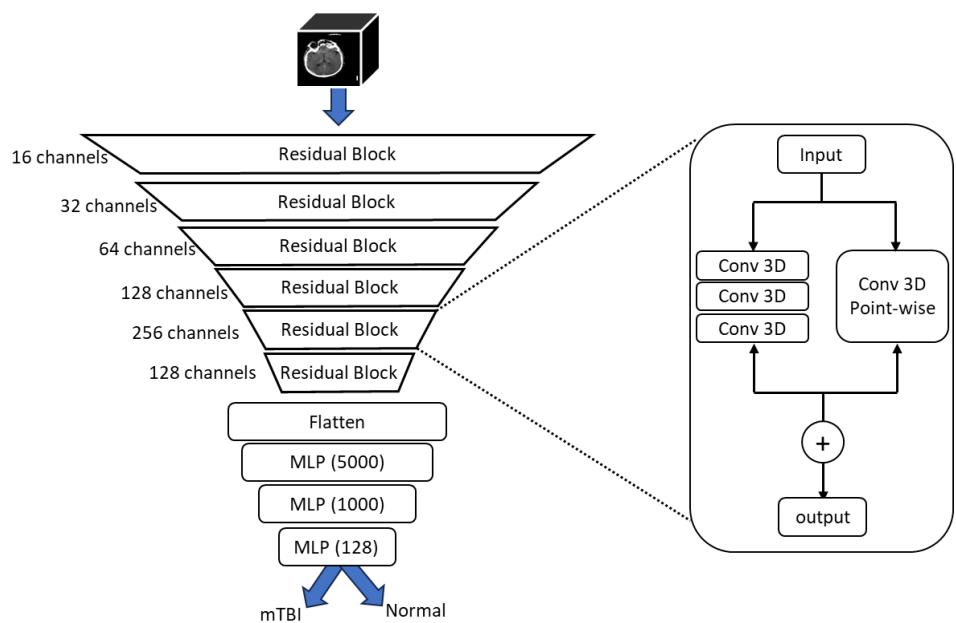

**Figure 2. RCNN model architecture**
MLP: Multi-Layer Perceptron

### 2.3.2 Residual Triplet Convolutional Neural Networks (RTCNN)

The RTCNN model is a powerful tool that leverages metric space embeddings for diagnosing mTBI using 3D images. This model employs the tuned residual CNN architecture , known for its ability to learn complex patterns and structures within data. The RCNN transforms 3D images into a high-dimensional embedding space, preserving the geometric relationships between the images.

Training our model involves the use of a Triplet Loss function, which operates on triplets of embeddings: an anchor (a randomly chosen embedding), a positive (an embedding of the same class as the anchor), and a negative (an embedding of a different class). The objective is to minimize the distance between the anchor and positive and maximize the distance between the anchor and negative in the embedding space. A key strategy in our approach is the concept of mining for hard positive and hard negative pairs. The hardest positive pair refers to the pair of embeddings from the same class that are farthest apart, while the hardest negative pair refers to the pair of embeddings from different classes that are closest together. By focusing on these challenging pairs, our model is encouraged to learn a robust metric space where similar classes cluster together, and different classes are distinctly separated.

The RTCNN model comprises four key components: an embedder, a miner, a sampler, and a classifier. These components work together to create a powerful and effective model for diagnosing mTBI, as illustrated in Figure 3. Moreover, a detailed description of the training and testing process can be found in Algorithm 1.

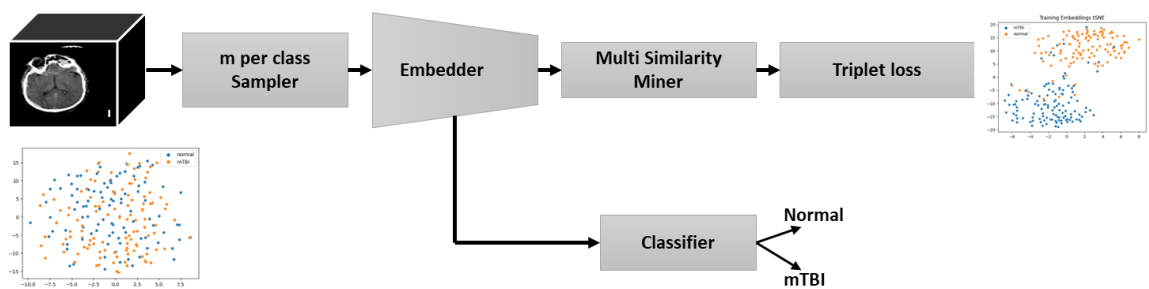

**Figure 3. Network architecture of RTCNN**

*2.3.2.1 Sampler*

The m per class sampler (MPCS) plays a crucial role in ensuring balanced batch formation with a specified number of samples from each class (m samples from both normal and mTBI), enhancing training control and potentially improving model performance. During each training loop iteration, the MPCS determines batch composition, aiming for 4 samples from each class due to m = 4. In the initial iteration, it randomly selects 4 samples from the normal class and 4 from the mTBI class, creating an 8-sample batch. To achieve the desired batch size of 32, this process is repeated 4 times, yielding 4 batches, each comprising 4 samples from the normal class and 4 from the mTBI class. Randomization within each class ensures sample diversity across batches.

*2.3.2.2   Embedder*

An embedder, in the context of metric learning, plays a fundamental role in transforming raw data, such as images, into a continuous and informative feature space. This feature space, often referred to as an "embedding space," is engineered to possess certain properties that facilitate tasks like similarity measurement and clustering. Essentially, the embedder encodes data points into a lower-dimensional space while ensuring that semantically similar points are placed closer together in this embedded space, simplifying tasks like classification or retrieval.

In our initial investigation into the influence of metric learning on diagnosis, we chose to utilize the RCNN encoder's architecture as the embedder for the RTCNN model. Additionally, we incorporated several dense layers with the aim of further compressing the embedded space vector. This enhancement was designed to empower the model in extracting and effectively leveraging informative features, ultimately contributing to the improvement of mTBI diagnosis. However, throughout our experimentation and hyperparameter tuning, we arrived at a simpler embedder configuration than the RCNN's encoder. Surprisingly, this simplified embedder not only proved effective but also outperformed the RCNN encoder. This unexpected outcome underscores the importance of rigorous experimentation and parameter tuning in optimizing model performance. Hence, we selected the finely tuned RCNN encoder to serve as the embedder for the RTCNN model.

The embedder consists of four residual blocks, with each block comprising three convolutional blocks and a 1×1 convolution layer. Within each convolutional block, there is a 3D convolutional layer, an instance normalization layer, and a parametric rectified linear unit (PReLU) activation function. Following the residual blocks, the output is flattened and linked to several dense layers, yielding a 512-dimensional image embedding vector. This architecture is depicted in Figure 4.

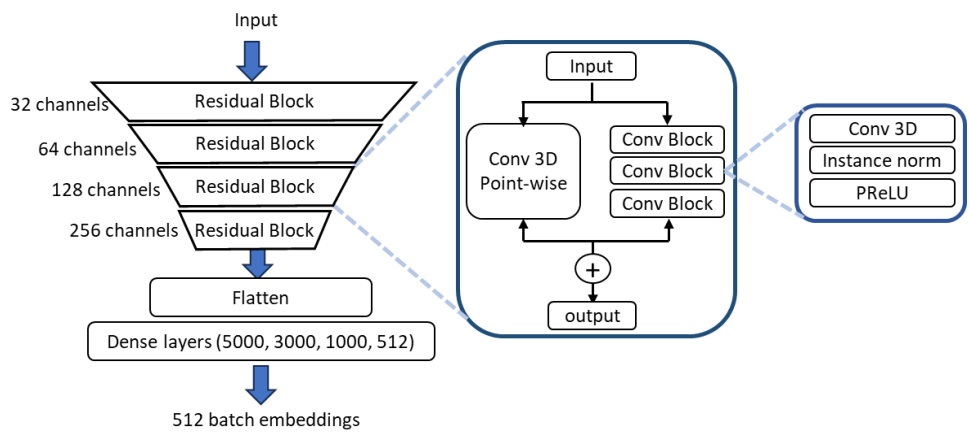

Figure 4. Embedder architecture of RTCNN

### 2.3.2.3 Miner

Miner is a strategic component designed to discover specific information within datasets/batches. Multi-Similarity Miner (MSM) receives a batch containing n embeddings and provide k pairs for subsequent loss calculation. Unlike traditional miners that may target a singular aspect or relationship, MSM extends its capabilities to capture diverse notions of similarity across various dimensions of the dataset [24]. The role of the MSM in our case is to select meaningful positive and negative pairs from the embeddings with a pre-set condition.

First, MSM identifies all positive pairs and negative pairs of the batch. Next, it assigns self-similarity weights for positive pairs and relative similarity weights for negative weights. For instance, if a mTBI image is very similar to another mTBI image in the batch (high self-similarity) but not similar to any normal images (low relative similarity), the positive pair of mTBI images might be assigned a high weight. Conversely, if a mTBI image is not very similar to another mTBI image but very similar to a normal image, the negative pair might be assigned a high weight. Subsequently, the selected pairs should satisfy the pre-set condition using a margin (epsilon=0.1), meaning selected negative pairs have similarity greater than the hardest positive pair minus epsilon, while the selected positive pairs have similarity less than the hardest negative pairs plus epsilon.

### 2.3.2.4 Triplet loss

The triplet loss is a central component of our model, enabling the model to acquire discriminative features. In the context of metric learning, the triplet loss is commonly expressed as:

$$L_{triplet} = \max \{0, d(a, p) - d(a, n) + \text{margin} \}.$$

This equation is crucial in training models to create embeddings where similar data points are brought closer together and dissimilar data points are separated by a defined margin. The following is the breakdown of the key elements:

- $d(a, p)$ represents the similarity between the anchor (a) and the positive example (p) in the embedding space.
- $d(a, n)$ signifies the similarity between the anchor (a) and the negative example (n) in the embedding space.
- 'margin' is a hyperparameter that's added to encourage a specific amount of separation between positive and negative pairs.

In practice, the loss is zero when the anchor-positive pair is closer to each other (i.e., $d(a, p)$ is smaller than $d(a, n)+\text{margin}$), which is the desired condition for similar pairs. Conversely, when this condition is not met, and the anchor-negative pair is closer (i.e., $d(a, n)+\text{margin}$ is smaller than $d(a, p)$), the loss becomes a positive value. In such a case, the model is encouraged to increase the distance between the anchor and the negative pair. Therefore, the triplet loss aims to ensure that the anchor is closer to the positive than to the negative by at least a margin. If this condition cannot be met, then the loss becomes positive, prompting adjustments to model parameters to minimize this loss. The similarity is calculated by the predetermined distance function. In our experiments, we configure the triplet loss to switch between the positive and anchor when the violation of the margin is more pronounced in the positive-negative relative similarity compared to the anchor-negative relative similarity.

The similarity between pairs is calculated based on the Euclidean distance or L2-norm to measure the distance between embeddings. The embeddings are normalized prior to distance calculation, which enhances model performance.

### 2.3.2.5 Classifier

The classifier is composed of three hidden layers with dimensions 256, 64, and 32, and it is capped by a sigmoid output layer. This architecture facilitates the learning of hierarchical features and efficient binary classification, taking the output from the trained embedder and distinguishing between mTBI and normal CT scans.

**Algorithm 1**: Learning Discriminative Embeddings with Residual Triplet Convolutional Neural Networks
**Input**: A dataset of images for classification.

**Output**: Trained Embedder, Trained Residual Triplet Convolutional Neural Network (RTCNN) Model, Visualized Embeddings, Occlusion Sensitivity Maps (OSM).

**Steps**:
1. Initialize the RTCNN model.
2. For each epoch, repeat the following steps:
3. Initialize m-Per-Class Sampler.
   - 3.1. Randomly sample "m" images from each class
   - 3.2. Continue sampling until reach the batch size of images.
4. For each batch of data from the sampler, perform the following:
   - 4.1. Compute embeddings for the batch using the Embedder.
   - 4.2. Calculate pairwise L2 distances/similarities between the embeddings.
   - 4.3. Apply the Multi-Similarity Miner to mine pairs where:
     - $d(a-n)_s > d(a-p)_h - \varepsilon$
     - $d(a-p)_s < d(a-n)_h + \varepsilon$
   - 4.4. Compute the triplet loss using all mined pairs.
     - $L_{triplet} = \max \{0, d(a, p)_s - d(a, n)_s + \text{margin} \}$
   - 4.5. Backpropagate the loss and update the parameters of the Embedder.
5. Return the trained Embedder and test it.
6. Visualize the embeddings using t-Distributed Stochastic Neighbour Embedding (t-SNE).
7. Freeze the Embedder parameters.
8. Connect the Embedder to a Multi-Layer Perceptron classifier.
9. Train the classifier using the discriminative embeddings.
10. Test the RTCNN model.
11. Generate the OSM.

## 3  Results and Discussion

Our model comprises two branches: the triplet loss branch and the classification branch. In this section, we provide and discuss the results obtained from each branch and how they contribute to the overall model performance.

### 3.1  Triplet Loss Branch and Embeddings

In the first branch, leveraging the triplet loss function, we successfully trained the model to produce highly discriminative embeddings. These embeddings encapsulate rich information about the underlying structure and relationships within the data. The key breakthrough in this branch lies in the quality of the embeddings. Through rigorous training, the model learned to arrange similar data points closely together while separating dissimilar ones in the embedding space. This not only provided a compact representation of the data, but also markedly improved the classification performance.

Upon visual inspection of the embeddings using t-distributed Stochastic Neighbour Embedding (t-SNE) [8], we observed clear clusters corresponding to each class, with minimal overlap between different classes. This observation highlights the effective minimization of intra-class distances and the maximization of inter-class distances achieved by the triplet loss function and mining process. Figure 5 depicts the embeddings for both the training and testing datasets, showcasing the transformations before and after the training procedure.

### 3.2  Classification Branch

The second branch of our model leverages these embeddings for classification, leading to notable improvements in classification accuracy and specificity. By enabling the model to distinguish between similar and dissimilar data points within the embedding space, the embeddings play a crucial role in capturing subtle yet critical patterns and features essential for accurate classification. As a result, the model achieved remarkable accuracy of 94.3%, effectively categorizing CT scans into mTBI and normal categories, as shown in Table 1. Table 1 presents the mean and standard deviation of accuracy, sensitivity, and specificity for the RTCNN, RCNN and Densenet121 models using 95% confidence intervals (CI).  Notably, the model exhibited a 22.5% improvement in specificity and a 11.3% increase in sensitivity compared to the RCNN model. Specificity, which measures the model's ability to accurately identify non-mTBI cases (true negatives) among all non-mTBI cases, reflects the model's ability in minimizing false positives and maintaining a high level of precision in identifying normal CT scans. Equally important, the model's heightened sensitivity signifies its improved ability to correctly identify mTBI cases (true positives) among all instances of mTBI. This balanced trade-off between accuracy, specificity, and sensitivity underscores the model's effectiveness as a tool for accurate mTBI diagnosis.

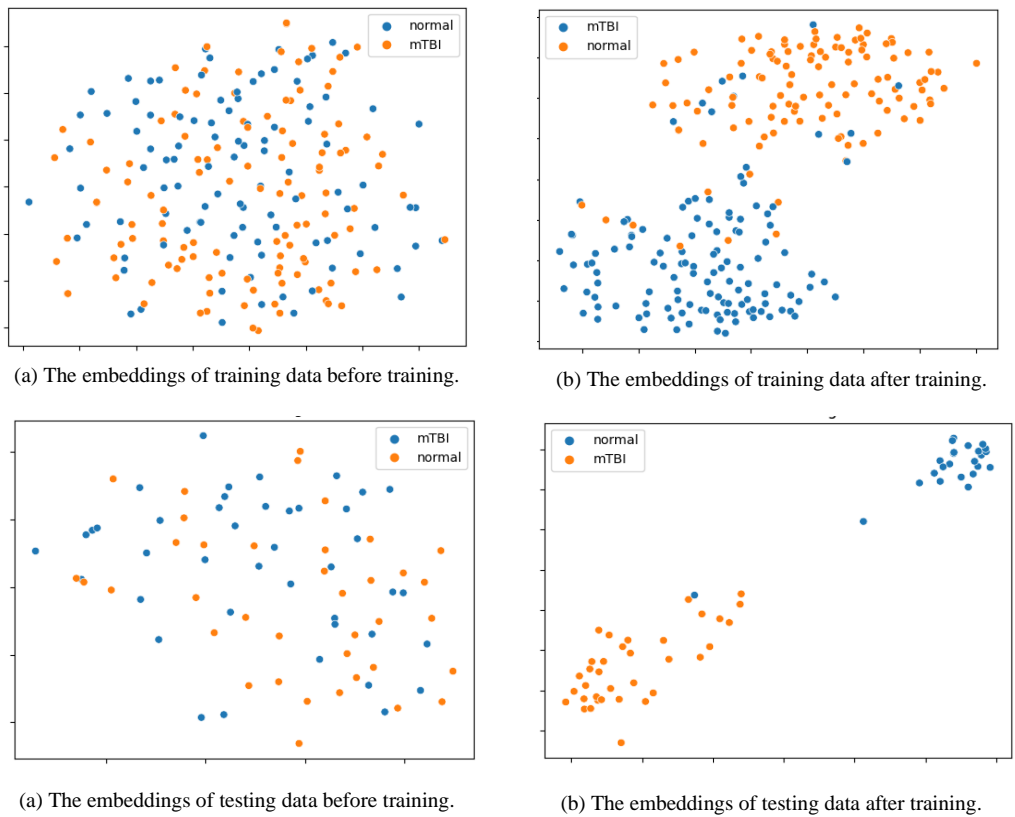

(a) The embeddings of training data before training.    (b) The embeddings of training data after training.

(a) The embeddings of testing data before training.    (b) The embeddings of testing data after training.

**Figure 5. the visual representation of the training and testing embeddings using t-SNE.**

Table 1: Densenet121, RCNN and RTCNN performance measures. The p-values provided are based on a one-sided t-test to establish a statistically significant improvement between RCNN and RTCNN.

| Metric | Accuracy ($\mu \pm \alpha$ in %) | Sensitivity ($\mu \pm \alpha$ in %) | Specificity ($\mu \pm \alpha$ in %) |
|---|---|---|---|
| Densenet121 [25] | 76.5 ± 2.6 | 77.5 ± 5.9 | 75.3 ± 5.3 |
| RCNN [10] | 78.1 ± 6.6 | 82.8 ± 5.1 | 72.7 ± 8.7 |
| RTCNN | **94.3 ± 4.0** ($p = 0.001$) | **94.1 ± 6.7** ($p = 0.020$) | **95.2 ± 3.4** ($p < 0.001$) |

Thus, the first branch of our model, guided by the Triplet Loss function and empowered by discriminative embeddings, significantly improved the accuracy, sensitivity, and specificity of mTBI classification. These achievements highlight the model's potential as a valuable asset in clinical decision support systems for diagnosing mild traumatic brain injuries.

Our proposed model not only achieves exceptional performance in terms of accuracy, specificity, and sensitivity but also distinguishes itself in efficiency when compared to other models, as demonstrated in Table 2. Despite having the highest number of parameters, our model does have superior memory efficiency during a forward/backward pass, resulting in significantly lower total memory requirements. This resource-efficient behaviour positions our model as a practical and suitable solution for real-world applications, where efficient memory utilization and parameter management are paramount for practical implementation. Therefore, our model not only delivers notable diagnostic capabilities, but does so efficiently, making it a compelling choice for mTBI diagnosis.

Table 2: Memory requirements for each implementation. MB is for megabytes.

|  | Number of Parameters | Forward/backward pass size (MB) | Params size (MB) | Total Memory Consumption (MB) |
|---|---|---|---|---|
| Densenet121 [25] | **11, 243, 649** | 1342.45 | **42.89** | 1389.34 |
| RCNN [10] | 35, 335, 490 | 1875.94 | 134.79 | 2014.73 |
| RTCNN | 66, 741, 964 | **448.58** | 254.60 | **707.18** |

## 3.3 Visual Assessment

The OSM is a powerful technique used to uncover the critical regions within an image that significantly impact a model's predictions. Through systematic occlusion of various parts of an input image and observing the resulting changes in the model's output, OSM reveals the model's focus and helps identify areas of both high and low importance [23]. This visual assessment using OSM offers an interpretable means to comprehend how the model arrives at its decisions, thereby facilitating model debugging, validation, and trustworthiness evaluation. In Figure 6, we present the OSM generated for a correctly classified mTBI CT scan by the RCNN and RTCNN models. These maps effectively elucidate the image regions that carry substantial weight in influencing the model's decision-making process. By distinguishing between areas of high (blue) and low (Red) importance, OSM provides valuable insights into the model's decision boundaries and helps identify crucial image landmarks, as well as potential sources of misclassification. Furthermore, it's noteworthy that the OSM of the RTCNN exhibits a broader distribution of important areas represented in blue. This observation indicates that the RTCNN model's decision-making process is influenced by a more extensive region within the image, suggesting a higher level of image context awareness. In contrast, the OSM generated by the RCNN model indicates a narrower focus on specific regions. This distinction underscores the RTCNN's exceptional capability to consider and integrate information from a larger portion of the input image, potentially contributing to its superior diagnostic performance and highlighting its advantage in understanding complex image data.

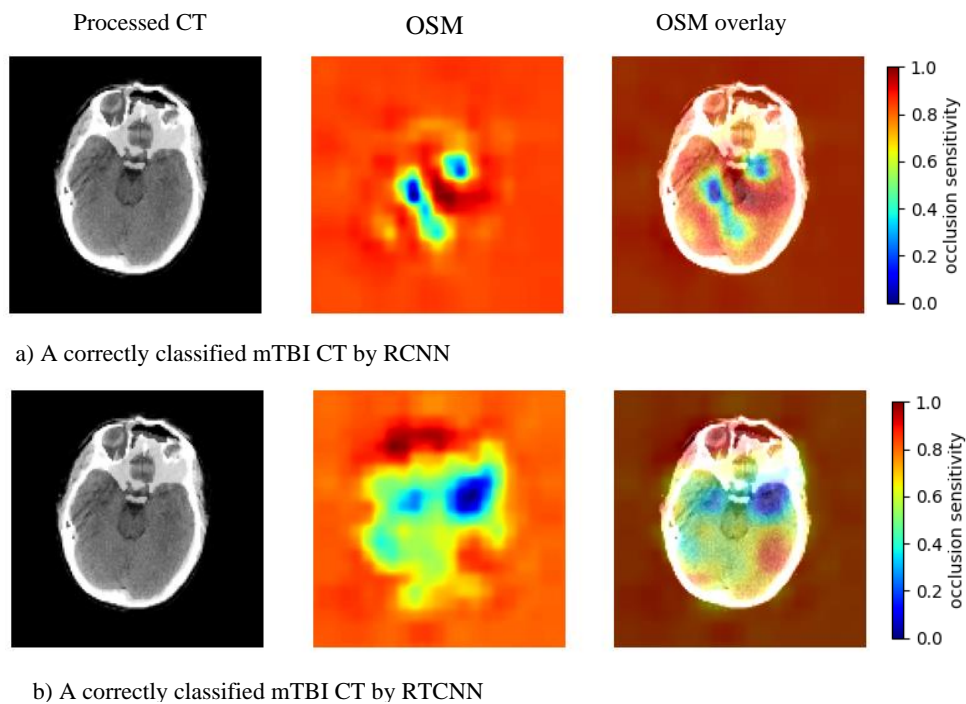

a) A correctly classified mTBI CT by RCNN

b) A correctly classified mTBI CT by RTCNN

**Figure 6. The Occlusion sensitivity map (OSM) for both the RTCNN and RCNN models.**
The blue areas represent the most critical regions that significantly influence the model's decision-making process.

## 4   Conclusion

In summary, our study represents a significant advancement in the challenging field of mTBI diagnosis. Our innovative approach, which integrates 3D Computed Tomography (CT) images with triplet loss-based metric learning to develop the Residual Triplet Convolutional Neural Network (RTCNN), can lead to substantial improvements in mTBI diagnosis. Compared to conventional methods, RTCNN demonstrates a remarkable 16.2% increase in accuracy, a substantial 22.5% improvement in specificity, and a notable 11.3% boost in sensitivity. Additionally, RTCNN has a lower memory consumption, making it not only highly effective but also resource-efficient in reducing false positives and accurately distinguishing between mTBI cases and normal CT scans. Furthermore, our use of occlusion sensitivity maps provide transparency for the model and form valuable insights into its decision-making process.

While our study has shown promising results in mTBI diagnosis, there are important limitations to acknowledge. The model's performance relies heavily on the availability of high-quality and precisely labelled CT scans for training. Variations in the quality and diversity of the training data can drastically affect the model's accuracy. Consequently, the collection and curation of such datasets pose substantial challenges that demand considerable effort. Future research endeavours should be dedicated to overcoming these limitations to further enhance the model's capabilities. One potential direction is the incorporation of clinical data to create a more comprehensive diagnostic tool. Furthermore, exploring alternative neural network architectures may lead to even better performance than what we have presented in this study. Addressing these issues will be essential in advancing the field of mTBI diagnosis and improving patient care.

## 5 Acknowledgement

NITRC, NITRC-IR, and NITRC-CE have been funded in whole or in part with Federal funds from the Department of Health and Human Services, National Institute of Biomedical Imaging and Bioengineering, the National Institute of Neurological Disorders and Stroke, under the following NIH grants: 1R43NS074540, 2R44NS074540, and 1U24EB023398 and previously GSA Contract No. GS-00F-0034P, Order Number HHSN268200100090U. Moreover, we would like to acknowledge the principal investigators of the TRACK TBI Pilot research program, the sub-investigators and research teams that contributed to TRACK TBI Pilot, and the patients who participated.

**Declarations of interest**: none